\documentclass[reprint,amsmath,amssymb,aps,superscriptaddress,prapplied]{revtex4-2}
\usepackage{graphicx}
\usepackage{amsmath}
\usepackage{gensymb}
\usepackage{amssymb}
\usepackage{multirow}
\newcommand{\cqt}{Centre for Quantum Technologies, National University of Singapore, 3 Science Drive 2, Singapore 117543, Singapore}
\newcommand{\ntu}{Division of Physics and Applied Physics, School of Physical and Mathematical Sciences, Nanyang Technological University, 21 Nanyang Link, Singapore 637371, Singapore}

\begin{document}


\title{Characterization of Josephson Junction Aging and Annealing Under Different Environments}

\author{Rangga P. Budoyo} 
\email{cqtrpb@nus.edu.sg}
\affiliation{\cqt{}}

\author{Rasanayagam S. Kajen} 
\affiliation{\cqt{}}

\author{Cheah Bing Wen} 
\affiliation{\ntu{}}

\author{Long H. Nguyen} 
\affiliation{\ntu{}}

\author{Rainer Dumke}
\email{rdumke@ntu.edu.sg}
\affiliation{\ntu{}}
\affiliation{\cqt{}}

\keywords{Josephson junction, junction aging, junction annealing, superconducting qubit}

\begin{abstract}
Understanding the aging behavior of Josephson junctions and the effect of annealing on junction resistances is important in building large-scale superconducting quantum processors. Here we study the effects of aging of Al/AlO$_x$/Al Josephson junctions under different storage conditions from immediately after fabrication up to 2 to 3 months. We find that the aging curve follows a logarithmic curve, with the aging amplitude mainly determined by fabrication conditions and the aging rate determined by both fabrication and storage conditions. Junctions stored at ambient laboratory conditions aged faster compared to junctions stored in a nitrogen atmosphere or vacuum, with the aging speed appreciably changes when the storage condition changed. We also compared the effect of thermal annealing under nitrogen environment with annealing under ambient conditions up to 250$\degree$C. We find that under nitrogen environment, the resistances decreased at all temperatures tested, while under ambient environment the resistances increased at 200$\degree$C and decreased at 250$\degree$C instead. We were unable to decrease the resistance below the initial-time resistance, suggesting a lower limit on the range of resistance tuning.
\end{abstract}
\maketitle

\section{Introduction}
\label{sec:intro}

Josephson junctions are essential elements in superconducting quantum processors~ \cite{Arute2019,Gao2025}. For all types of superconducting qubits, their frequencies depend on the critical current $I_c$ of the junctions, which in turn depends on fabrication conditions including the junction area and oxidation parameters. At room temperature, $I_c$ can be estimated by measuring the junction resistance $R$, which is related to $I_c$ via the Ambegaokar-Baratoff relation~\cite{Ambegaokar1963}. For large-scale quantum processors, precise and accurate junction fabrication is important for qubit frequency assignment to avoid frequency collisions that can limit qubit gate performance~\cite{Hertzberg2021, Morvan2022}. Current state-of-the-art fabrication performance gives about 3\% wafer-scale $I_c$ coefficient-of-variation (CV) and 1-2\% chip-scale $I_c$ CV~\cite{Kreikebaum_2020, Takahashi2022, Zheng2023}.

To improve the frequency accuracy, several junction anneal methods that change the critical current post-fabrication are developed. These include thermal annealing (i.e heating), which has been reported to decrease $I_c$ (increase $R$) or increase $I_c$ (decrease $R$) depending on the temperature and the heating environment~\cite{Scherer2001, Koppinen2007, Korshakov2024}. However, thermal annealing is limited to chip-scale adjustment only, while junction-scale annealing will be necessary for frequency accuracy improvement. In recent years, laser annealing~\cite{Hertzberg2021, Zhang2022, Kim2022, Du2026}, electron-beam annealing~\cite{Balaji2024}, and voltage annealing \cite{Pappas2024, Wang2024, Kennedy2025a} are developed that allows junction-scale tuning. Junction-scale tuning were previously limited to only decreasing $I_c$, although there are recent reports where laser annealing increased $I_c$ instead~\cite{Du2026}. 

An additional factor that needs to be taken into account is the fact that the junctions ``aged", where the resistance slowly increased over time~\cite{Koppinen2007, Nesbitt2007}. As a result, a qubit's measured frequency can differ significantly from its expected frequency if its $R$ is measured significantly before it is cooled down. Recent reports show that junction aging is affected by junction design and fabrication, including junction dimensions and oxidation parameters~\cite{Krizan2026}. Several methods have been shown to inhibit aging, including plasma cleaning~\cite{Pop2012}, additional oxidation steps~\cite{Bilmes2021}, or using etching instead of lift-off process~\cite{VanDamme2024}. However, reducing the aging appears to also reduce the junction sensitivity to annealing~\cite{Krizan2026, Bilmes2021}, suggesting a balance between the two is needed in the manufacturing of large scale quantum processors.

In this work, we study the Al/AlO$_x$/Al junction aging properties under different storage environments over several months, including ambient atmosphere, nitrogen glove box, and vacuum. 
We observed that the aging amplitude appears to be limited mainly by fabrication conditions, similar to recent reports~\cite{Krizan2026}. 
On the other hand, aging rate/timescale appears to both depend on the fabrication and storage conditions, with ambient conditions resulted in fastest aging. 
Changing the storage conditions of the chips results in an apparent change in aging rate, with a resistance decrease ($\equiv$ junction ``deaging") observed when the junctions are moved from ambient atmosphere to glove box. 
We also studied the effect of prior storage on both voltage annealing and thermal annealing of Josephson junctions. The results from voltage annealing experiments suggest this process did not induce an accelerated aging of junctions, but instead changed the internal structure of the junctions. For thermal annealing in nitrogen atmosphere, we only observed decrease in $R$ up to 250$\degree$ C temperatures, consistent with previous reports in low-oxygen environment~\cite{Scherer2001, Korshakov2024}. For thermal annealing in ambient conditions, we observed increase in $R$ for 200$\degree$ C and decrease in $R$ for 250$\degree$ C. Additionally, the lower limit for resistance decrease appeared to be the initial (immediately after fabrication junction) $R$ value. Understanding junction aging and annealing characteristics allows one to plan the optimal storage and the timing and annealing method for different resistance targets.

\section{Test Junction and Storage Details}
\label{sec:expe}

The test Al/AlO$_x$/Al junctions used in this work were fabricated using the Dolan junction fabrication procedure~\cite{Dolan1977} widely used in the fabrication of superconducting qubits~\cite{Moskalev2023,Muthusubramanian2024}. Each test chip consists of 16 junctions, with a typical fabricated size of approximately 250 nm $\times$ 250 nm. On each chip there are also 4 shorted test structures with similar contact pad design to verify probe-pad contact reliability.

High-resistivity silicon wafers ($> 10000 \ \Omega$ cm) are diced into individual 1.5 cm $\times$ 1.5 cm square chips. We spin-coated bilayer e-beam resist, consisting of 500 nm MMA-EL9 and 300 nm PMMA-A4-950K. After drawing the test junction patterns using e-beam lithography, the chips are developed using MIBK:IPA 1:3 solution for 40 s, and then loaded into the e-beam deposition system. 

Before deposition, we performed in-situ cleaning and SiO$_2$ removal using a gentle argon ion mill (beam voltage 250 V)~\cite{Quintana2014} at $+60\degree$ and $0\degree$ tilt angles. A small flow of oxygen (Ar flow 9 sccm, O$_2$ flow 0.9 sccm) is included to remove  organic residue. Two layers of Al are deposited with 0.5 nm/s rate, with different thicknesses and deposition tilt angles: 1st layer 60 nm and $+60\degree$, and 2nd layer 100 nm and $0\degree$. We performed an oxidation step between the deposition of the 2 layers, with 0.5 mbar oxygen pressure and 6 minutes duration. After the second deposition, we performed a final protective oxidation step for 10 mbar and 10 minutes. We performed liftoff using heated acetone (about 50$\degree$C) for approximately 45 minutes. This liftoff duration was shorter than the typical duration (several hours or overnight) so that we can measure the junction resistance as close to the deposition time as possible. We also did not perform additional bandage layer fabrication~\cite{Dunsworth2017}. As a result, the chips were not exposed to further baking step that may result in any significant thermal annealing. 
For this work, we fabricated 6 chips in 3 separate depositions separated by several days, see Table~\ref{tab:chiplist}. For each deposition, 2 chips are deposited simultaneously. All depositions used the same Al crucible, with very similar chamber pressures and deposition currents.

To measure the resistance of the test junctions, we use a 2-probe measurement using a semi-automated probe station. Using a source measure unit, we sweep the applied current between -25 nA and +25 nA while measuring the voltage across the junction, with 51 points per sweep. The resistance is given by the slope of the $I$--$V$ curve. For each junction at each time we performed two current sweeps, and the reported individual junction resistances are the average from the two sweeps. 
The initial resistance measurement are typically done within 20 minutes after the end of lift-off (corresponding to about 4 hours after oxidation step), then the chips are stored in their assigned environment. 

\begin{table*}[htp]
    \centering    
    \caption{Summary of test chips reported in this work.}
    \label{tab:chiplist}
    \begin{tabular}{c c c c c}
        \hline
         Fabrication & Chip & $\langle R(t\approx0)\rangle$ & Storage & Anneal \\
         & &  (k$\Omega$) & & \\
         \hline
         \hline
         \multirow{2}{*}{1} & 1 & $22.8\pm4.8\%$ & Ambient Atmosphere & Voltage (Half) \\
         & 2 & $24.3\pm5.9\%$ & Nitrogen glove box & Voltage (Half)\\
         \hline
        \multirow{2}{*}{2} & 3 & $7.5\pm3.9\%$ & Atmosphere $\leftrightarrow$ glove box, alternating with Chip 4 & Thermal \\
         & 4 & $8.7\pm5.1\%$ & Glove box $\leftrightarrow$ atmosphere, alternating with Chip 3 & Thermal \\
         \hline
        \multirow{2}{*}{3} & 5 & $11.1\pm3.3\%$ & Vacuum for 7 days, then glove box  & None \\
         & 6 & $11.1\pm12.1\%$ & Nitrogen glove box & None \\
         \hline
    \end{tabular}
\end{table*}

The chips are stored in a gel box, either in ambient atmosphere on a cupboard shelf in a (non-clean room) laboratory room, with typical temperature of 22-23$\degree$ C and relative humidity of about 60\%, or inside a continuously purged nitrogen glove box made from acrylic, with $(0.0\pm0.1)\%$  measured O$_2$ population, typical temperature of 19-20$\degree$ C, and $(1\pm1)\%$ measured relative humidity. Assuming an upper limit of $0.1\%$ O$_2$ concentration and $1\%$ humidity, we have an upper limit of 1000 ppm and about 300 ppm for O$_2$ and H$_2$O concentrations, respectively. These values are comparable to the typical O$_2$ and H$_2$O concentration in acrylic glove boxes.
Chip 5 was temporarily stored inside the high-vacuum space of the e-beam deposition chamber used in the junction fabrication with a base pressure of $\sim1\times10^{-7}$~mbar. 
As the probe station was located in ambient lab environment, during the resistance measurements the chips would need to be taken out of their storage for approximately 20 minutes.  For 5 of the 6 chips, the initial resistance CV are between 3 and 6\%. For Chip 6, the higher CV is because of a lower fabrication yield, with a region of open junctions. Table~\ref{tab:chiplist} shows a summary of the chips reported in this work.

\section{Comparison of Aging Between Different Environments}
\label{sec:aging}

\begin{table*}[htp]
    \centering    
    \caption{Summary of the fractional aging $\langle \Delta R/R(t\approx0)\rangle$ of all chips at several chosen ages. The underlined values mean the chip was swapped to a different environment after the measurement while the empty cells mean the chip was not measured at that specific time. For single-environment chips and Chip 5, the average of fit parameters of the aging to the logarithmic curve of Eq.~\ref{eq:logfit} is also shown.}
    \label{tab:fitpars}
    \begin{tabular}{c c c c c c c}
        \hline
         $t$ & \multicolumn{6}{c}{Chip} \\
          (days) & 1 & 2 & 3 & 4 & 5 & 6 \\
        \hline
        \hline
        1 & $(44\pm2)\%$ & $(18\pm2)\%$ & $\underline{(32\pm3)\%}$ & $\underline{(12\pm1)\%}$ & $(11\pm1)\%$ & $(13\pm1)\%$\\
        2 & $(57\pm3)\%$ & $(27\pm2)\%$ & $(35\pm3)\%$ & $(34\pm2)\%$ & $(16\pm1)\%$ & $(17\pm1)\%$ \\
        4 & $(70\pm4)\%$ & $(34\pm3)\%$ & $\underline{(39\pm3)\%}$ & $\underline{(53\pm3)\%}$ & $(22\pm2)\%$ & $(24\pm2)\%$ \\
        5 & - & - & $(52\pm4)\%$ & $(52\pm3)\%$ & - & - \\
        7 & $(79\pm6)\%$ & $(40\pm3)\%$ & $\underline{(60\pm4)\%}$ & $\underline{(54\pm4)\%}$ & $\underline{(27\pm2)\%}$ & $(30\pm3)\%$ \\
        8 & - & - & $(60\pm5)\%$ & $(65\pm4)\%$ & $(38\pm3)\%$ & $(32\pm3)\%$ \\
        13 & - & - & $\underline{(64\pm6)\%}$ & $\underline{(81\pm7)\%}$ & - & - \\
        14 & $(93\pm9)\%$ & $(48\pm5)\%$ & $(70\pm6)\%$ & $(80\pm7)\%$  & $(45\pm4)\%$ & $(37\pm3)\%$ \\
        21 & $(106\pm11)\%$ & $(58\pm5)\%$ & $(84\pm7)\%$ & $(79\pm7)\%$ & $(50\pm5)\%$ & $(41\pm5)\%$  \\
        End & $(126\pm16)\%$ & $(76\pm8)\%$ & $(98\pm8)\%$ &  $(82\pm7)\%$ & $(57\pm6)\%$ & $(49\pm7)\%$  \\
        & (56 days) & (56 days) & (85 days) & (88 days) & (46 days) & (46 days) \\
        \hline
        $\langle a\rangle$ & $0.21\pm0.05$ & $0.17\pm0.03$ & - & - & $0.12\pm0.02$ & $0.10\pm0.02$  \\
        $\langle\tau\rangle$ ($10^4$ s)&  $1.4\pm0.8$ & $5.2\pm2.2$ & - & - & $6.1\pm2.6$ & $3.2\pm1.6$\\
        $\langle b\rangle$ & $0.80\pm0.22$ & $0.98\pm0.06$ & - & - & $0.92\pm0.03$ & $0.78\pm0.14$\\
        \hline
        Fit range & 56 days & 56 days  & - & - & First 7 days & 46 days  \\
        \hline
        $\langle a\rangle$ & - & - & - & - & $0.12\pm0.03$ & -  \\
        $\langle\tau\rangle$ ($10^4$ s)&  - & - & - & - & $3.1\pm1.7$ & -\\
        $\langle b\rangle$ & - & - & - & - & $0.89\pm0.17$ & -\\
        \hline
        Fit range & - & -  & - & - & Day 8 to 46 & -  \\
        \hline
    \end{tabular}
\end{table*}

\subsection{Single Environment Chips}
\label{subsec:single}

\begin{figure*}[tpb]
    \centering
    \def\svgwidth{\columnwidth}
    \includegraphics[width=1.9\columnwidth]{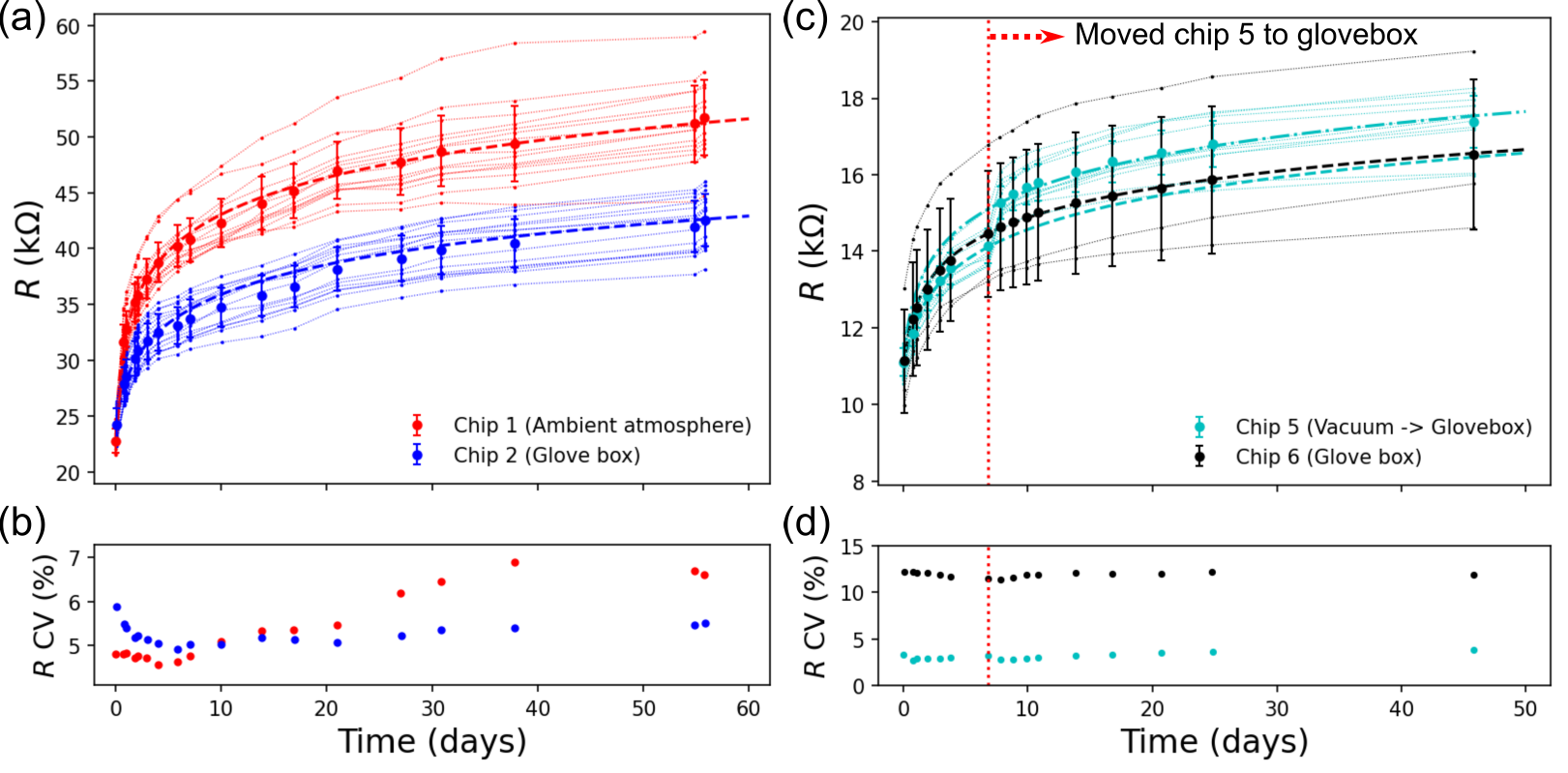}
    \caption{(a,c)~Junction resistance $R$ as a function of time $t$ after junction fabrication. Small dots correspond to the resistances of each junction, with their uncertainties typically smaller than the dot size. The circle symbols and their error bars correspond to the the average and the standard deviation of the resistance. The dashed and dashed-dotted lines correspond to the aging curve of Eq.~\ref{eq:logfit} using the average of fit parameters of Table~\ref{tab:fitpars}. (b,d)~Resistance coefficient of variation (CV) as a function of time $t$ after junction fabrication. (a,b)~Chips 1 and 2 were fabricated at the same time, then stored in ambient atmosphere (Chip 1) or nitrogen glove box (Chip 2). (c,d)~Chips 5 and 6 were fabricated at the same time, then initially stored in vacuum (Chip 5) or nitrogen glove box (Chip 6). At $t=7$ days, Chip 5 was moved from vacuum to glove box (vertical red dotted line). }
    \label{fig:chip12}
\end{figure*}

The resistances of junctions on Chip 1 (stored in ambient atmosphere) and Chip 2 (stored in glove box), which were simultaneously deposited, were measured over a period of 56 days. Fig.~\ref{fig:chip12}(a) shows the individual and averaged resistance of the test junctions during the measurement period, with $t=0$ defined as approximately the time the chips are taken out from the acetone bath at the end of lift-off. The uncertainties of the individual resistance measurement are typically smaller than the dot size. Fig.~\ref{fig:chip12}(b) show the resistance CV of the resistances over the measurement period. For Chip 1 it grew slowly to about $7\%$, while for Chip 2 it remained about the same.

We fit the resistance aging for each junction to a phenomenological logarithmic fit \cite{Kennedy2025a,Krizan2026}
\begin{equation}
    \frac{R(t)}{R(t\approx0)} = 1 +  a \ \log \left(\frac{t}{\tau}+b\right)
    \label{eq:logfit}
\end{equation}
where $a$ corresponds to the fractional aging amplitude, $\tau$ corresponds to aging characteristic timescale. We include an offset parameter $b$ inside the logarithm to ensure the value remains finite at $t=0$, and we expect $b\approx1$ since the argument of the logarithm should approach unity at early times.
Table~\ref{tab:fitpars} shows the fit parameters for Chips 1 and 2, averaged over all the junctions on each chip (see the Supplementary Materials for the histogram of the fit parameters). The parameter $b$ were $\approx1$ for both chips, consistent with expected value. The aging amplitude $a$ for both chips are quite similar, with overlapping distribution, while aging timescale $\tau$ for Chip 2 are about 3.5 times higher than the $\tau$ for Chip 1.

Figure~\ref{fig:chip12}(c) shows the resistance $R$ of Chips 5 and 6, which were also deposited simultaneously as a function of time $t$, with the coefficient of variation (CV) of resistance shown in Fig.~\ref{fig:chip12}(d). Chip 5 was initially stored in vacuum, then at $t=7$ day (vertical dotted line in Fig.~\ref{fig:chip12}) it was moved to glovebox. Chip 6 was stored in glovebox during the entire 46 days of measurement. 

In Table~\ref{tab:fitpars} we also included the average of fit parameters obtained from the junctions in Chip 6 and the first 7 days of Chip 5 (while stored in vacuum). $b$ for Chips 5 and 6 are also $\approx 1$. The parameter $a$ for Chips 5 and 6 are similar, and smaller than $a$ for Chips 1 and 2, and $\tau$ for Chip 5 is about 2 times larger than for Chip 6. 
Thus, we can infer that the aging amplitude (corresponding to $a$) mainly depend on the fabrication condition, similar to recent reports~\cite{Krizan2026}. Meanwhile, the aging timescale (corresponding to $\tau$) depends on the storage condition. However, the different value of $\tau$ between Chips 2 and 6 suggests fabrication also contributes to aging timescale.

\subsection{Alternating Environment Chips}
\label{subsec:alternating}

\begin{figure}[tpb]
    \centering
    \def\svgwidth{\columnwidth}
    \includegraphics[width=.95\columnwidth]{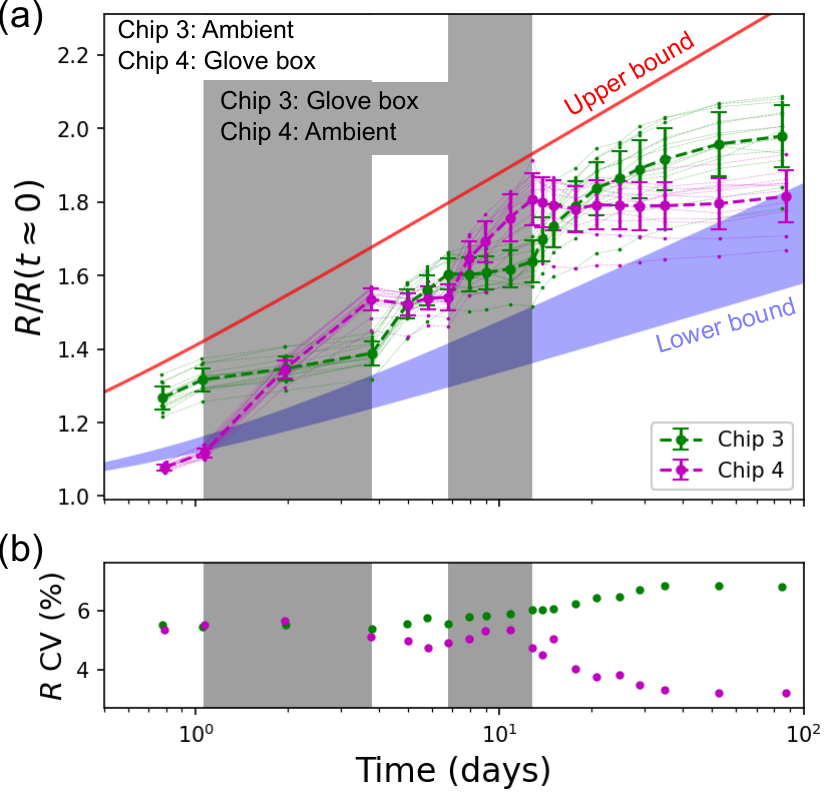}
    \caption{(a)~Fractional resistance aging $R(t)/R(t\approx0)$ for Chips 3 and 4 as a function of time $t$ after junction fabrication (in semilogarithmic $x$-axis). Small dots correspond to the values for each junction. The circle symbols and their error bars correspond to the average $\langle R(t)/R(t\approx0)\rangle$ and the standard deviation of the fractional resistance. Chip 3 was stored in glove box while Chip 4 was stored in atmosphere during the shaded period, and vice versa during the non-shaded period. The red line and blue region correspond to the estimated aging bounds for the average $\langle R(t)/R(t\approx0)\rangle$ based on the aging of Chips 1, 2 and 6. (b)~Resistance coefficient of variation (CV) as a function of time $t$ after junction fabrication.}
    \label{fig:chip345}
\end{figure}

Chips 3 and 4 were deposited simultaneously, and Chip 3 was initially stored in atmosphere while Chip 4 was initially stored in glove box. Their junction resistances were measured over a period of about 85 days, with their storage condition swapped several times during this period, as shown in Fig.~\ref{fig:chip345}(a), with Fig.~\ref{fig:chip345}(b) showing the resistance CV. For Chip 3, CV increased slowly, especially after the last sample swap to atmosphere. Considering the increase in CV we observed for Chip 1, this would be expected. The decrease of CV for Chip 4 after $t>12$ days appears to be partially caused by the shorting of one of the junctions during the measurement period.

As the chips are swapped, the apparent aging rate (i.e. the local gradient of the aging curve) changed significantly (see also Table~\ref{tab:fitpars}). When the chips were moved from glove box to atmosphere, the chips aged more rapidly. When the chips were moved from atmosphere into glove box instead, the aging slowed significantly. When this swapping happens after $t\approx4$~days, we observed some junctions' resistance decreased ("deaging") instead. After the last swapping at $t=13$~days, we observed no apparent aging for more than 40 days, with average resistance change of only about 1\% during this period.

The fractional aging of Chips 3 and 4 appears to be bounded by the expected aging fully in atmosphere and fully in glove box. These bounds are shown as the red line and blue regions in Fig.~\ref{fig:chip345}(a) and based on the aging curve of Chips 1, 2, and 6. As we showed in the previous section, the amplitude of the aging varied depending on the chip/fabrication and so the exact bound can vary between chips. The relaxation of the resistance towards these bounds appear to be quite slow, with apparent decay time of order several days.

The aging curves for Chips 3 and 4 does not appear to follow a single logarithmic curve of Eq.~\ref{eq:logfit}. Instead, for each storage period, except for the initial one, phenomenologically it appears to follow
\begin{equation}
    \frac{R(t)}{R(t\approx0)} = 1 +  a \ \log \left(\frac{t}{\tau}+b\right) \left[1+c\,e^{-(t-t_m)/\tau_e}\right],
    \label{eq:2logfit}
\end{equation}
the logarithmic aging of Eq.~\ref{eq:logfit} defining the aging bounds, with an exponential decay towards these bounds. $c$ is the exponential component amplitude, with $c<0$ when the chips were moved from glove box to air, and vice versa. $t_m$ is the time the chip were moved between environments, and $\tau_e$ the characteristic exponential decay time. However, considering the limited number of measurement points for each period and the 5 fit parameters ($a$, $b$, $\tau$, $c$, $\tau_e$) in Eq.~\ref{eq:2logfit}, it is difficult to obtain a reliable fit for both chips for each period. However, from Fig.~\ref{fig:chip345} it appears $\tau_e$ depends on the swapping time, with an apparent relaxation of order $\sim$days after the first environment swapping at $t=1$ day, and of order $\sim$weeks after the last swapping at $t=13$ days, i.e. $\tau_e$ is comparable to junction age.

In contrast to the gradual relaxation observed in Chips 3 and 4, there was a rapid increase ($\tau_e\ll1$ day) in resistance when Chip 5 was moved from vacuum to glove box [see Table~\ref{tab:fitpars} for $t=7$ and 8 days and Fig.~\ref{fig:chip12}(c)]. This suggests the relaxation time $\tau_e$ towards the aging bound depends on the conditions (oxygen level, humidity) of the previous and current environment. Due to this fast relaxation, and assuming the same initial $R(t=0)$, the resistance of Chip 5 while in the glovebox can be fitted using the single logarithmic function of Eq.~\ref{eq:logfit}, and shown also in Table \ref{tab:fitpars}. While $a$ and $b$ were relatively unchanged, $\tau$ is reduced by a factor of 2, and is comparable to the value of $\tau$ for Chip 6 which was also stored in the glovebox.
\section{Junction Annealing}
\label{sec:anneal}

\subsection{Alternating Bias Voltage Annealing}
\label{subsec:abaa}

\begin{table*}[htp]
    \centering
    \caption{Average resistance change and aging fit parameters after junction voltage annealing.}
    \label{tab:vanneal}
    \begin{tabular}{c c c c c c c}
          \\
          \hline
          Chip & Junction &  $\langle\Delta R\rangle$  & $\langle\Delta R/R\rangle$  & $\langle a'\rangle$ & $\langle a' R_0\rangle$  & $\langle \tau_e\rangle$  \\
          & Set & (k$\Omega$) & (\%) & (\%) & (k$\Omega$) & ($10^5$ s) \\
         \hline
         \hline
         1 & Annealed & $7.4\pm1.0$ & $14.2\pm1.0$ & $4.3\pm2.5$ & $2.6\pm1.8$ & $2.7\pm2.1$ \\
         & Non-Annealed & $0.07\pm0.07$ & $0.1\pm0.1$ & - & - & -  \\
        \hline
         2 & Annealed & $7.6\pm0.4$ & $18.1\pm1.4$ & $3.8\pm0.7$ & $1.9\pm0.4$ & $1.4\pm0.4$ \\
         & Non-Annealed & $0.17\pm0.06$ & $0.4\pm0.2$ & $1.8\pm0.4$ & $0.8\pm0.2$ & $1.6\pm1.1$ \\
         \hline
    \end{tabular}

\end{table*}

\begin{figure}[tpb]
    \centering
    \def\svgwidth{\columnwidth}
    \includegraphics[width=.95\columnwidth]{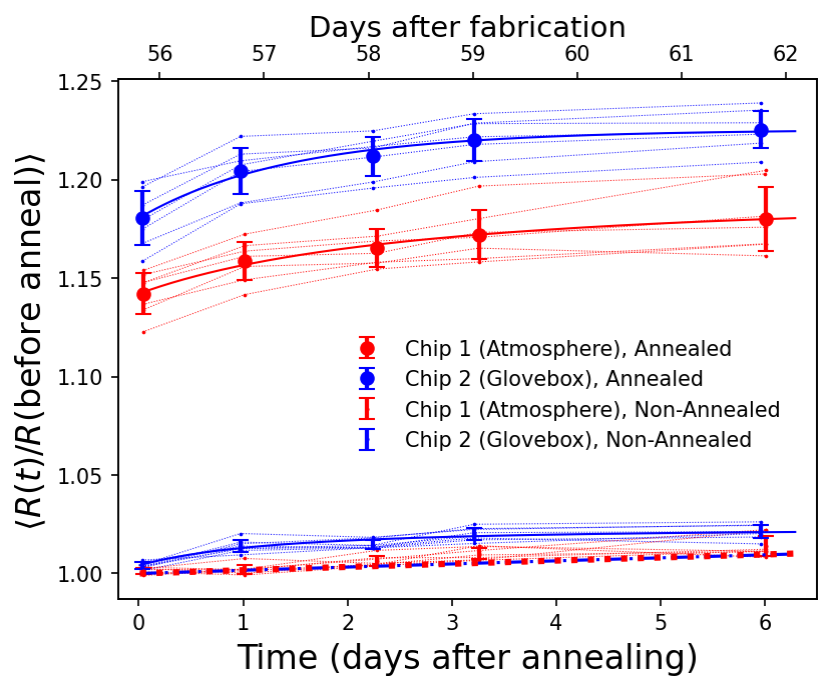}
    \caption{Fractional resistance change of the test junctions as a function of time after the voltage annealing process. The thin dotted lines correspond to the fractional resistance of each junction. The solid circle symbols and their error bars correspond to the average and standard deviation of fractional resistance for each chip. The thick red dotted lines and blue dashed lines correspond to the expected aging of the non-annealed junctions on Chips 1 and 2, respectively, extrapolated based on the fit parameters in Table~\ref{tab:fitpars}. The solid lines correspond to the fit of the post-anneal aging to the exponential function of Eq.~\ref{eq:expfit}}
    \label{fig:chipanneal}
\end{figure}

After the aging measurements described in Section \ref{subsec:single}, at $t=56$ days after junction deposition, we annealed half of the junctions on Chips 1 and 2 using the alternating-bias assisted annealing (ABAA) method~\cite{Pappas2024, Wang2024}.
We applied a sequence of 30 alternating voltage (15 positive, 15 negative) square pulses with $\pm 0.9$ V amplitude and 1 s duration. A heating element allows us to heat the chips up to about $75\degree$ C, but for the junctions reported in this work, all the voltage annealing were done at room temperature.

Table~\ref{tab:vanneal} summarizes the effect of annealing process on the junctions. The non-annealed junctions were not affected by the annealing of nearby junctions. The annealed junctions on both chips increased by an average of approximately 7.5 k$\Omega$, which corresponds to about $14\%$ resistance increase for Chip 1 and about $18\%$ increase for Chip 2.

If we assume voltage annealing as "rapidly aging" the junctions, we can then define the "added age" from the annealing process, and this added age should be similar for both chips as they were under the same annealing process. 
For Chip 1, we can estimate the post-anneal junction age by using the fit parameters from Table~\ref{tab:fitpars} to calculate $t$ from Eq.~\ref{eq:logfit} for the measured post-anneal resistance. For Chip 1, we estimated $t\approx300$ days for an added age of 244 days.
For Chip 2, since the annealing is done at ambient environment, this estimate needs to be adjusted. Based on the results of Section~\ref{subsec:alternating}, one would expect the added aging to follow Eq.~\ref{eq:2logfit}. Instead, we assume that this added aging causes the resistance curve to reach a single-logarithmic bound curve of Eq.~\ref{eq:logfit} with $a\approx0.17$, the original value of $a$ for Chip 2, but with $\tau\approx1.4\times10^4$ s, which is the $\tau$ value of Chip 1 while in ambient conditions. This gives a post-anneal age of $t\approx76$ days. This value is significantly less than the post-anneal age for Chip 1 above, and as a result one likely should not consider voltage annealing as rapid aging. Instead, with the similar initial post-fabrication $R(t\approx0)$ and the post-annealing resistance change $\Delta R$, the voltage annealing likely changed the internal structures of the junction. 


Afterwards, we measured the post-annealed resistances for another 6 days, as shown in Fig.~\ref{fig:chipanneal}. During these 6 days, Chip 1 was again stored in ambient atmosphere, and Chip 2 was stored in glovebox.
The annealed junctions on both chips follow a similar resistance curve, with initial aging before slowing down after several days, similar to previous reports \cite{Wang2024, Krizan2026}. This post-anneal aging would be detrimental to the accuracy of qubit frequency targeting if the qubits were not immediately cooled down after annealing.
For the non-annealed junctions on Chip 1, their resistances follow the expected resistance (red dotted line, extrapolated to $t\gtrsim56$ days based on fit parameters in Table~\ref{tab:fitpars}), while there was a small increase for Chip 2. This is likely due to the fact that during the annealing process, the chip was out in atmosphere for approximately 90 minutes, 4 to 5 times longer than the typical time it was in the atmosphere during the aging measurements in the previous section, and was under an accelerated aging for a short period.
 
Similar to the swapping experiment in Section~\ref{subsec:alternating}, the expected aging curve would be the logarithmic-exponential curve of Eq.~\ref{eq:2logfit}, with the logarithmic term defining the resistance bound, and the exponential term defining the post-anneal relaxation towards this bound. However, we only tracked this aging for a relatively short time for aged junctions, where the average resistance of non-annealed junctions in Chip 1 only increased by $(1.4\pm0.5)\%$ after 6 days. As a result, the logarithmic term can be approximated as a constant value.

Instead, we fit the resistance aging to the exponential function
\begin{equation}
    \frac{R(t)}{R_0} = 1 +  a' \left(1-e^{-t/\tau_e}\right),
    \label{eq:expfit}
\end{equation}
with $t=0$ redefined as the voltage annealing time (and $t_m=0$), $R(t\approx0)\equiv R_0$ redefined as the resistance immediately after annealing, and $c=1$.  The average of the fit parameters are shown in Table~\ref{tab:vanneal}, with the fit lines shown as solid lines in Fig.~\ref{fig:chipanneal}. The exponential aging amplitude $\langle a' \rangle$ (and $\langle a R_0\rangle$, in units of k$\Omega$) for annealed junctions in Chip 2 is approximately twice the $\langle a' \rangle$ for non-annealed junctions in Chip 2. This means that for the annealed junctions on Chip 2, there is an additional aging from annealing in addition to the change in environmental conditions (as shown by the non-annealed junctions), which is also consistent with the results for Chip 1.



\subsection{Thermal Annealing}
\label{subsec:thermal}

\begin{table*}[htp]
    \centering
    \caption{Thermal annealing steps performed on Chips 3 and 4. Wait time corresponds to the duration the chip is left in the oven after the end of annealing before the resistance measurement. The fractional resistance change $\Delta R/R$ were scaled to the resistance measurement prior to that anneal step (after the previous step). }
    \label{tab:tanneal}
    \begin{tabular}{c  c  c  c  r  r}
        \\
        \hline
          Step & Temperature  &  Environment & Wait time & \multicolumn{2}{c}{$\langle \Delta R/R\rangle$ ($\%$)}  \\
          & ($\degree$C) & & (minutes) & Chip 3 & Chip 4 \\
         \hline
         \hline
         1 & 200 & N$_2$ & 10 &  $-4.0\pm1.8$ & $-8.8\pm1.6$\\
         2 & 250 & N$_2$ & 10 &  $-27.2\pm1.1$ & $-36.1\pm1.0$ \\
         3 & 200 & ambient & 40 &  $9.1\pm1.0$ & $9.8\pm0.7$ \\
         4 & 250 & ambient & 10 &  $-17.5\pm1.6$ & $-6.2\pm0.7$\\
         5 & 200 & ambient & 10 & $1.9\pm1.4$ & $2.6\pm0.7$ \\
         \hline 
         - &  & aging for 119/116 days & & $14.2\pm4.1$ & $9.0\pm6.7$  \\
         & & & & \multicolumn{2}{c}{(Chip 3 anomalies not included.)} \\
         \hline
    \end{tabular}

\end{table*}

\begin{figure}[tpb]
    \centering
    \def\svgwidth{\columnwidth}
    \includegraphics[width=.95\columnwidth]{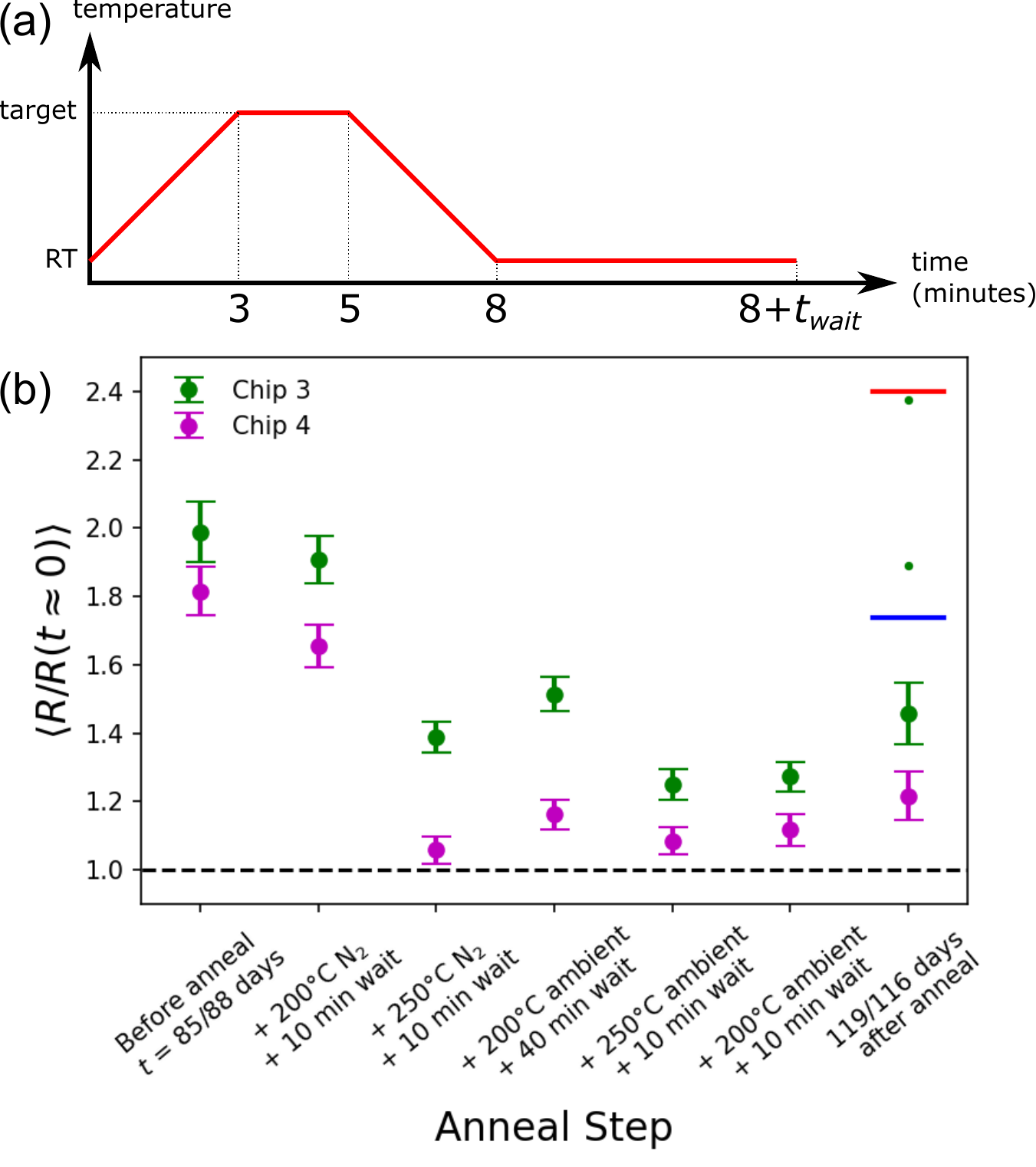}
    \caption{(a) Heating curve of the reflow oven during the thermal annealing measurement. After a wait time $t_\text{wait}$ (typically 10 minutes), the chips were removed from the oven, then their junction resistance measured. (b)Average fractional resistance of the test junctions after each thermal annealing process. The horizontal dashed line corresponds to the $t\approx0$ resistance value. For the measurements 119/116 days (for Chip 3/4, respectively) after the anneal, the 2 outliers on Chip 3 (green dots) were not included in the averaging. The red and blue line correspond to the upper and lower bounds for the resistance, respectively, based on the bounds in Fig.~\ref{fig:chip345} extrapolated to $t=117.5$ days.}
    \label{fig:tanneal}
\end{figure}

After the aging measurements described in Section~\ref{subsec:alternating}, at $t=85/88$ days (for Chips 3/4, respectively), we performed thermal annealing on Chip 3 (stored in ambient atmosphere for about 70 days) and Chip 4 (stored in glove box for the same amount of time). 
The thermal annealing was performed using a reflow oven with a maximum temperature of 280$\degree$ C, either in a nitrogen environment ($<1\%$ measured O$_2$ population just outside the oven during the process) or in an ambient environment ($\approx21\%$ O$_2$ population). The anneal process takes 8 minutes, with 3 minutes heating time to the specified temperature, 2 minutes hold time, and 3 minutes of cooling time [see Fig.~\ref{fig:tanneal}(a)]. Afterwards, the chip was left in the oven for some time (typically about 10 minutes) for thermalization before the resistance was measured.

The annealing for each chip was done at different days, but follow the exact same sequence shown in Table~\ref{tab:tanneal}. After each step, we measured the junction resistances, shown in Fig.~\ref{fig:tanneal}(b). From the results, we can infer the following:
\begin{enumerate}
    \item We observed no difference in the general annealing behavior (resistance increase / decrease) between the chips, although the magnitude of the fractional resistance change was significantly larger for Chip 4 for annealing in N$_2$.
    \item For annealing in N$_2$, the resistances decreased for the temperatures used, with a larger decrease for 250$\degree$ C. This is consistent with previous reports for thermal annealing in low-oxygen environment, either in forming gas (H$_2$/N$_2$ mixture) \cite{Scherer2001} or in argon \cite{Korshakov2024}.
    \item For annealing in ambient atmosphere, we observed a resistance increase for annealing at 200$\degree$ C, and a resistance decrease at $250\degree$ C. The decrease at $250\degree$ C was smaller than the decrease for the same temperature in N$_2$ atmosphere. The decrease at higher temperatures appear to show a reverse effect compared to thermal annealing at low-oxygen environments, where $R$ increase emerged at high temperatures 400$\degree$ C and above instead \cite{Scherer2001, Korshakov2024}.
    \item Even though steps 3 and 5 have the exact same annealing process, we observed a larger resistance increase for step 3. This is likely due to the natural junction aging caused a further increase during the additional 30 minutes of thermalization.
    \item Here we performed sequential thermal annealing steps on aged junctions. The resistance change may differ when only a single annealing step was performed on each chip, or when annealing freshly fabricated junctions. However, there was previous evidence showing similar response from laser annealing between fresh and aged junctions~\cite{Kim2022}. 
    \item We were unable to decrease the resistance to below the initial value, i.e. we could not deage the junctions to a $t<0$ time. Thus, the minimum tuning range of junction resistance appeared to be set by the initial resistance. Whether this would be a general lower resistance limit for thermal annealing would require further study.
\end{enumerate}

After the thermal annealing, Chip 3 was again stored in ambient conditions and Chip 4 was stored in glove box. 119/116 days after the thermal annealing process (for Chips 3/4, respectively), we remeasured the resistance of the junctions on both chips, where all junctions aged further. 
If the resistance decrease from thermal annealing resulted in the reduction of the effective age of the junctions (to near $t\approx0$), we can estimate the expected aging using the bounds in Fig.~\ref{fig:chip345}. In Fig.~\ref{fig:tanneal}, the red and blue lines show these bounds extrapolated to $t=117.5$ days. For most junctions on Chip 3 and all junctions on Chip 4, their aging was slower than this estimate, suggesting that the annealing does not actually reduce the effective age of the junctions.
Two of the junctions in Chip 3 aged to within the bounds instead. Understanding the aging difference between the junctions would require further study.

\section{Discussion}
\label{sec:discussions}

In this work, we reported the difference in the aging of Josephson junctions in different storage conditions, as well as voltage annealing and thermal annealing of some of those junctions. Junction aging was thought to be due to adsorption of oxygen~\cite{Shiota1992} or desorption of hydrates (OH)~\cite{Konkin1979} to/from the AlO$_x$ layer, where the latter may be influenced by the presence resist residues~\cite{Pop2012}, and junction annealing was thought to cause an artificial acceleration of the aging process~\cite{Koppinen2007}.

In chips stored in a single environment during its lifetime (Section~\ref{subsec:single}), their resistances followed a logarithmic dependence. Logarithmic functions have previously been used to describe the stress/strain relaxation of polymers, colloids, or glassy systems~\cite{Lomnitz1956, Guiu1964, Amir2012}. The emergence of logarithmic decay in such systems was thought to be due to a distribution of exponential relaxation rates in the system \cite{Amir2012}.
Glassy behavior has previously been reported in Al/AlO$_x$/Al junctions, with their resistance following an aging trajectory that depends on applied bias voltage, and a relaxation towards this curve when the voltage bias is changed~\cite{Nesbitt2007}. 
Junction aging can then be considered as an effect of the relaxation of glassy AlO$_x$, which consists of parallel resistors/conductors, each with their own conductance $G_i=1/R_i$ and  relaxation rate $\Gamma_i$. Considering Eq.~\ref{eq:logfit} gives $R(t\to\infty)=\infty$, we can then assume each conductor relaxes towards $G_i(t\to\infty)\approx0$ equilibrium.
It is then reasonable to expect that the resistance aging follows a logarithmic function similar to Eq.~\ref{eq:logfit}, which was what we observed. 

Similar to \cite{Nesbitt2007}, we also expect different environmental control parameters to result in different aging trajectories (i.e. Eq.~\ref{eq:logfit} fit parameters). Taking into account the likely sources of aging discussed above, we can then consider the following: 
\begin{itemize}
    \item The aging amplitude $a$ can be considered as a measure of the difference between the initial and long-time ($G_i\to0$) AlO$_x$ configurations. The specific long-time configuration is determined by the initial (post-deposition) configuration, among other things. (e.g. the equilibrium can be a local minimum near the initial point in the potential landscape).
    \item The aging rate $1/\tau$ (i.e. average of $\Gamma_i$ distribution) depends on the oxygen adsorption and hydrate desorption rate, which in turn depends on the concentration of oxygen and water in the storage environment, as well as the presence of the resist residue (i.e. from the fabrication process). 
\end{itemize}
This agrees with the observation that $a$ depends on the fabrication conditions (e.g different fabrication batches) and $\tau$ depends on both fabrication conditions and storage conditions.

When junctions are moved to a different storage condition, their baseline aging curve would shift to a different Eq.~\ref{eq:logfit} trajectory based on the new storage. As changing storage would not induce an immediate structural changes on the AlO$_x$, their instantaneous resistance would be outside this aging trajectory. As a result, the resistance curve would relax towards the new trajectory. This relaxation can be modeled as an exponential decay with the moving time $t_m$ as the initial time. Thus, it is reasonable to model the aging curve in the swapping experiment (Section~\ref{subsec:alternating}) as logarithmic-exponential function of Eq.~\ref{eq:2logfit}.

The growth and subsequent self-limiting of native AlO$_x$ on an Al surface is typically modeled using Cabrera-Mott mechanism~\cite{Cabrera1949}. In this model, Al$^{3+}$ ions on the metal-oxide interface and O$^{2-}$ ions on the oxide surface generate a potential difference across the oxide (called the Mott potential), resulting in an electric field across the dielectric. This allows electrons to tunnel through the oxide to create further ionization, allowing the oxide to grow until the field is reduced as the thickness increases. During the voltage annealing (Section~\ref{subsec:abaa}), the applied electric field across the junction oxide can facilitate further oxide growth, resulting in increase in resistance~\cite{Pappas2024}. This will also result in a new structural configuration post-annealing with a new equilibrium state as well. 
Since previous results suggest that voltage bias or laser annealing does not change the effective age of junctions~\cite{Kim2022, Nesbitt2007}, the post-anneal aging can be described using Eq.~\ref{eq:2logfit}, with the logarithmic function describing the single-environment aging trajectory, and the exponential function describing the relaxation towards this trajectory.
In our case, we tracked the post-anneal aging of aged junctions for a relatively short period of time (for 6 days from $t=56$ days). During this period, from Eq.~\ref{eq:logfit} and Table~\ref{tab:fitpars} we expect $\langle R\rangle$ for Chip 1  to increase by 0.49 k$\Omega$ ($0.95\%$), and $\langle R\rangle$ for Chip 2 to increase by 0.41 k$\Omega$ ($0.95\%$).
As a result, we can approximate the logarithmic term as a constant value $a \log[(t/\tau)+b]\equiv\gamma$. Then, we have
\begin{equation}
    R(t) \approx R(t=0) \left\{1 +  \gamma  \left[1+c\,e^{-(t-t_m)/\tau_e}\right] \right\},
    \label{eq:2logapprox}
\end{equation}
with $c<0$ in the voltage annealing case.
If we redefine the time $t'=t-t_m$ where $t_m$ here is the junction age during annealing, we return to the exponential relaxation starting at $t'=0$ similar to Eq.~\ref{eq:expfit}, which we used to fit the post-annealing aging in Section~\ref{subsec:abaa},
\begin{equation}
    \frac{R(t')}{R(t'=0)} = 1 +  a' \left(1-e^{-t'/\tau_e}\right),
    \label{eq:2logapproxexp}
\end{equation}
with $R(t'=0) = [1+\gamma(1+c)] R(t=0)$ and $a'=-\gamma c/[1+\gamma(1+c)]$. 

In Section ~\ref{subsec:thermal}, thermal annealing in N$_2$ environment resulted in a resistance decrease for both 200$\degree$C and 250$\degree$C temperatures.
This resistance decrease suggests the reversal of processes that cause aging, i.e. desorption of oxygen and absorption of hydrates instead. 
Thermal annealing in ambient environment resulted in a resistance increase at 200$\degree$ C and a decrease at 250$\degree$ C that is smaller than in N$_2$. This suggests that the presence of oxygen or moisture in the ambient environment caused an oxygen adsorption/hydrate desorption process that counteracts the oxygen desorption/hydrate adsorption during annealing. The inability to reduce $R$ below its initial value suggests a limit where these processes are suppressed, up to the  "as-fabricated" configuration. However, this observation was based on the results of a single batch of fabrication (Chips 3 and 4), under sequential thermal annealing steps. It would be interesting to study further the range of anneal parameters (i.e. temperature, duration, environment, junction age, single/sequential anneal) that result in the emergence of the resistance lower limit. 

The structural changes in the junction oxide due to aging and annealing can be verified and further studied by imaging the cross-section of the junction, for example by using a scanning tunneling electron microscope (STEM), with the chemical composition extracted using energy-dispersive X-ray spectroscopy (EDX) or electron-energy loss spectroscopy (EELS) \cite{ Korshakov2024, Du2026}.

\section{Conclusions and Outlook}
\label{sec:conclusion}

In summary, we observed that the logarithmic aging curve of Josephson junction resistances depends on both fabrication conditions and storage conditions. In our case, the difference in oxidation conditions mainly changes the aging amplitude. On the other hand, aging rate depends on both storage and fabrication conditions. Swapping between storage conditions show appreciable change in apparent aging rate, with a small deaging when moving from ambient conditions to glove box. 

We also compared the effect of voltage and thermal annealing on junctions with identical fabrication but different storage conditions. The similar $R$ change and further aging post-voltage annealing suggest that voltage annealing does not ``accelerate" the environmental aging process, but instead changed the internal configuration of the junctions. We observed resistance decrease for thermal annealing in N$_2$ environment for all temperatures used (up to 250$\degree$ C). For thermal annealing in ambient conditions, we observed a slight increase at 200$\degree$ C and decrease at 250$\degree$ C, suggesting a competition between deaging and aging (e.g oxygen desorbption/absorption) processes during annealing.

Even though the resistances were measured in ambient conditions, we expect the added environmental effects during the measurement to be minimal in the aging curve. This is because each measurement time ($\approx$ 20 minutes for a full chip) is much shorter than the time between each measurement (days, up to 1 month). Nevertheless, it might be interesting to compare our results with chips fully stored and measured in the same environment. Additionally, there are additional storage environments of interest, for example: 
\begin{enumerate}
    \item purified closed-loop stainless steel glove boxes with 0.1-1 ppm O$_2$ and H$_2$O population,
    \item dehumidifying dry cabinets with low ($<20\%$) relative humidity and near-normal O$_2$ population, and
    \item oxygen-enriched environments with $>21\%$ O$_2$ population.
\end{enumerate}

Our results show that to control junction aging, particular care should be taken in junction storage conditions. Uncontrolled storage between qubit junction resistance measurement and cooldown can result in unexpected qubit frequency shift beyond the desired range. While high-vacuum storage resulted in the slowest aging, the large resistance jump we observed after moving the sample out of the vacuum chamber will result in  undesired qubit frequency jump. Instead, the most optimal storage condition appears to be N$_2$ glove box, with comparable aging speed.

\acknowledgments{The authors thank Senthil Kumar Karuppannan, Lilis Rachel Yohana, Christoph Hufnagel, and Yuanzheng Paul Tan for their assistance and useful discussions, and D. Scott Holmes and Tim Smith for valuable suggestions.

This project is supported by the National Research Foundation, Singapore through the National Quantum Office, hosted in A*STAR, under its Centre for Quantum Technologies Funding Initiative (S24Q2d0009) and by the National Research Fund (NRF) Quantum Engineering Program (QEP) W21Qpd0211 Advanced Quantum Processor Platform.}
\bibliography{main}

\end{document}